\documentclass{article}
\usepackage[preprint]{spconf}

\usepackage{amsmath,graphicx,color}
\usepackage[T1]{fontenc} % add special characters (e.g., umlaute)
\usepackage[utf8]{inputenc} % set utf-8 as default input encoding
\usepackage{cite,url}
\usepackage{multirow}
\usepackage{booktabs}
\usepackage[bookmarks=false,hidelinks]{hyperref}

\def\mycopyrightnotice{%
	\begin{minipage}{\textwidth}
		\scriptsize
		\copyright~2023 IEEE. Personal use of this material is permitted. Permission from IEEE must be obtained for all other uses, in any current or future media, including reprinting/republishing this material for advertising or promotional purposes, creating new collective works, for resale or redistribution to servers or lists, or reuse of any copyrighted component of this work in other works.
	\end{minipage}
}

\title{Tempo Estimation as fully self-supervised Binary Classification}
\name{Florian Henkel, Jaehun Kim, Matthew C. McCallum, Samuel E. Sandberg, Matthew E. P. Davies}%\thanks{Thanks to XYZ agency for funding.}}
\address{SiriusXM-Pandora, USA}
\begin{document}
\ninept
\maketitle

\begin{abstract}
	This paper addresses the problem of global tempo estimation in musical audio. Given that annotating tempo is time-consuming and requires certain musical expertise, few publicly available data sources exist to train machine learning models for this task. Towards alleviating this issue, we propose a fully self-supervised approach that does not rely on any human labeled data.  Our method builds on the fact that generic (music) audio embeddings already encode a variety of properties, including information about tempo, making them easily adaptable for downstream tasks. 
	While recent work in self-supervised tempo estimation aimed to learn a tempo specific representation that was subsequently used to train a supervised classifier,
	we reformulate the task into the binary classification problem of predicting whether a target track has the same or a different tempo compared to a reference. While the former still requires labeled training data for the final classification model, our approach uses arbitrary unlabeled music data in combination with time-stretching for model training as well as a small set of synthetically created reference samples for predicting the final tempo. Evaluation of our approach in comparison with the state-of-the-art reveals highly competitive performance when the constraint of finding the precise tempo octave is relaxed.
\end{abstract}


\copyrightnotice{\mycopyrightnotice}
\section{Motivation}
\label{sec:intro}
Incorporating machine learning techniques, in particular in the form of deep neural networks (DNNs), has yielded significant improvements in music tempo estimation \cite{schreiber18ismir,boeck20ismir}. However, DNNs tend to require large amounts of (often human annotated) data to perform and generalize well. For example in the case of tempo estimation, having access to diverse data that covers a wide range of genres can further boost the prediction performance of these networks \cite{schreiber18ismir}.
Given that curating large annotated data sources is costly and time consuming, self-supervised approaches to tempo estimation have recently emerged in the scientific literature \cite{quinton22ismir, morais2023tempo}.
Inspired by these works,
we propose to tackle tempo estimation in a fully self-supervised way by reformulating this task as a binary classification problem -- does track A have the same or a different tempo compared to a reference track B. 

The advantage of such a formulation is that we do not require any labeled data for model training and only need a small set of reference tracks (one or more examples for each tempo we want to predict). 
Moreover, we show that this reference set can consist entirely of synthetic, i.e., artificially and automatically created, tracks, in which case, our contribution represents the first fully self-supervised approach to tempo estimation to be quantitatively evaluated on real data.  

\begin{figure*}[ht!]
	\centerline{\includegraphics[width=0.9\textwidth]
		{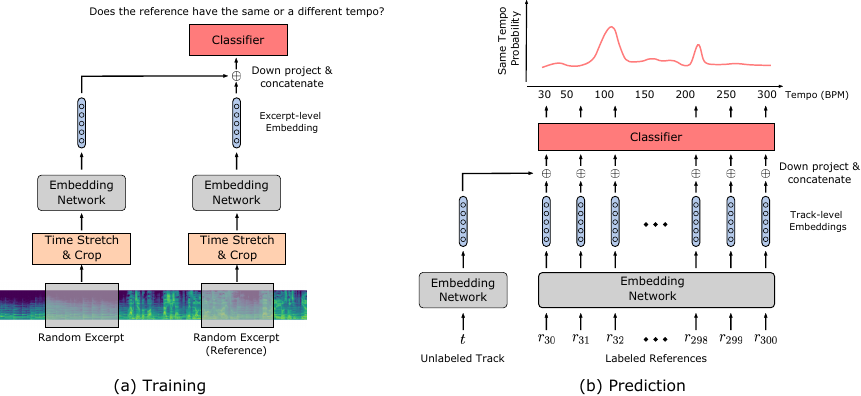}}
	\caption{
		Overview on the task setup. For training (a), 
		we sample random mel-spectrogram excerpts and the task is to predict whether they still have the same tempo after being time-stretched (with potentially different stretch factors) and cropped to comprise $3$ seconds of audio.
		For prediction (b) we compare an unlabeled track against all reference tracks and look for the reference with the highest probability for \emph{same tempo} which is then reported as the predicted tempo.
		Note that the embedding network is fixed, i.e., not updated during training, and the same in all stages.
	}
	\label{fig:task_setup}
\end{figure*}

\section{Related Work}
\label{sec:related_work}

Tempo estimation is a fundamental task in music information retrieval (MIR) with a wide range of applications including, but not limited to, automatic playlist generation or music search and retrieval (for a broader overview on applications refer to \cite{schreiber2020music}).

While early approaches to tempo estimation primarily relied on hand-crafted features, heuristics and domain-knowledge \cite{gouyon2006experimental,mckinney07jnmr}, DNNs are now ubiquitous.
In \cite{schreiber18ismir}, the authors propose to directly predict tempo using convolutional neural networks based on mel-spectrogram inputs.
Likewise in \cite{boeck20ismir}, tempo estimation is considered jointly alongside beat and downbeat estimation in a multi-task setup using temporal convolutional networks. 
More recently, \cite{quinton22ismir} proposes to learn audio embeddings that are specifically suited for tempo estimation using self-supervised learning with an equivariance constraint. These embeddings are subsequently used to train a tempo classifier in a supervised fashion with labeled data. Thus, the actual tempo classification itself is not fully self-supervised.

A closely related approach for self-supervised tempo estimation is proposed in \cite{morais2023tempo}. In contrast to the equivariance constraint method in \cite{quinton22ismir}, tempo estimation is addressed as a reformulation of SPICE for pitch estimation \cite{gfeller2020spice}. The primary goal is to identify the offset between vertically shifted tempogram slices \cite{grosche11taslp} together with a secondary goal of reconstructing the tempogram slice. This is followed by a calibration stage using synthetic data in the form of isolated click tracks over a range of tempi to map the predicted offset to an absolute tempo value. While this approach does not require any labelled data in the form of real-world musical content with tempo annotations, and thus mirrors our approach in this respect, it is formulated as a regression problem whereas we pose the task as binary classification. In addition, the evaluation focuses on the properties of the synthetic data distributions from which the training data is sampled and the subsequent impact on the calibration, and contains no quantitative evaluation of tempo estimation performance on real-world annotated datasets.

\section{Methodology}
\label{sec:method}

In the following, we first provide details on how we use music audio embeddings in this work and subsequently we describe our approach, looking at the training and prediction stages separately. An overview of both stages is shown in Figure \ref{fig:task_setup}.

\subsection{Music Audio Embeddings}

Instead of learning an audio representation from scratch as in \cite{quinton22ismir}, we make use of existing generic music audio embeddings in the form of MULE \cite{mccallum22ismir}. The intuition behind using these kinds of embeddings is that all the `heavy lifting' has already been performed by a task-independent model and that the resulting embedding or representation contains pertinent information for various downstream tasks.
In particular it has been shown that such latent representations can already implicitly encode tempo information making them suitable for tasks such as beat tracking \cite{li2023mert} or tempo prediction \cite{mccallum24icassp}. Furthermore, \cite{mccallum24icassp2} demonstrates the manipulation of tempo directly in the MULE embedding space.  

Using contrastive learning \cite{chen2020simclr}, MULE is trained in a self-supervised fashion by sampling random pairs of mel-spectrogram excerpts within a temporal neighborhood of $10$ seconds of the same track. 
These mel-spectrogram excerpts (comprising $3$ seconds of audio) are encoded into $1728$-dimensional embeddings, which can subsequently be used in two forms.
First, \emph{as-is}, or what we refer to as \emph{excerpt-level} embeddings, and second, in the form of \emph{track-level} embeddings by averaging all of the \emph{excerpt-level} embeddings of a track along its time-line.

In the following, we use \emph{excerpt-level} embeddings for model training and \emph{track-level} embeddings for prediction.
It has been found that both variants can achieve similar results when used for downstream tasks \cite{mccallum22ismir}. Hence our assumption is that training a classifier based on excerpts also generalizes to the track-level, making the prediction more efficient.

\subsection{Training}

Unlike most recent approaches to tempo estimation, our goal in this work is not to train a classifier to directly predict tempo as a single beats-per-minute (BPM) value,
but rather to predict whether one track has the same or a different tempo compared to a reference.
Given that we use unlabeled data for training, we do not have (nor do we need) any information on how the tempi of the two tracks relate to one another. However, we explicitly leverage the assumption that for the task of global tempo estimation, we should expect the tempo within a given track to be largely constant. On this basis, we consider musical content which is highly expressive in terms of tempo variation to be outside the scope of this work.

Sampling two excerpts from the same track (at different temporal locations) allows us to time stretch 
the excerpts and then simply create a label for this training pair depending on whether the stretching results in a tempo change or not as determined by the sampled stretch-factors, which can be the same for both excerpts. 
To allow for more efficient model training, we do not perform time stretching on the audio waveform, but directly augment the (pre-computed) mel-spectrograms using a cubic spline interpolation similar to \cite{schreiber18ismir}. 
In sum, both excerpts comprise $3$ seconds of audio after time stretching and cropping, and are encoded to $1728$-dimensional embeddings using MULE. These embeddings are down projected, concatenated and used as input to a shallow neural network classifier that is trained using binary cross-entropy loss with the aforementioned automatically generated label of ``same'' or ``different''. In our experiments we jointly train a single shared down projection layer to further reduce the dimensionality of the embeddings.

\subsection{Prediction}
For prediction, we rely on a small number of reference examples to determine the actual BPM of an unlabeled input track.
For each BPM tempo we want to predict, we require at least one example, i.e., if we want to predict tempo in a range of 30 to 300 BPM, we need a reference track with tempo 30, 31, 32, etc. 
Using MULE we compute the (down projected) track-level embeddings for all references $r_{30}$ to $r_{300}$  as well as for the unlabeled track $t$ for which we want to know the tempo. In a next step, we concatenate the embedding of $t$ with each reference, apply the classifier and predict whether $t$ has the \emph{same} or a \emph{different} tempo as the reference.\footnote{In case of multiple references for each tempo one could simply average the output logits or the probabilities after applying a sigmoid function.
}
Given these predictions, one way to arrive at a final tempo is to simply take the \emph{argmax} of the \emph{same-tempo} probability across all references. 
In the following we will refer to our same-or-different approach as SDNet\textsubscript{argmax}.

\section{Experimental Setup}

For our experiments, we train a shallow classifier with two dense layers of size $128$, a shared down projection layer of size $256$, ReLU activation function \cite{nair2010rectified}, a batch size of $256$ and Adam optimizer \cite{kingma14adam}. To convert the model outputs to probabilities we apply a sigmoid function.
The model is trained over $20$k steps with an initial learning rate of $0.001$ which is annealed to $0$ over time following a cosine learning rate scheduler with $2000$ warm-up steps \cite{LoshchilovH17_SGDR_ICLR}.
For time stretching during training we randomly sample stretch factors in a range of $[0.75, 1.5]$ following a log-uniform distribution. This range is selected to avoid artifacts which are potentially introduced by interpolating the mel-spectrogram using more extreme stretch factors.
To ensure that the model is exposed to an equal number of samples for both the \emph{same} and \emph{different} category, we set the same stretch factor for the excerpts $50$\% of the time.

\subsection{Data and Evaluation Metrics}
For model training we use an arbitrary set of $1.7$M musical tracks. As we do not require any labels, these tracks could come from any source (e.g., AudioSet \cite{gemmeke2017audioset}
or the Million Song Dataset \cite{bertinMahieux2011msd}), however in our particular case we leverage a small subset of music from the catalog of a large music streaming service.
For the reference tracks we restrict ourselves to a purely synthetic setup and thus avoid the introduction of any human labelling of tempo into our pipeline. Specifically, we create MIDI tracks with different tempi in the range of 30 to 300 BPM consisting of a sequence of C4 quarter notes. These tracks are rendered to audio waveforms with a piano soundfont using Fluidsynth.\footnote{\url{https://www.fluidsynth.org/}}
For the parameters of the mel-spectrogram computation we follow \cite{mccallum22ismir}.
As is common in the literature \cite{schreiber18ismir, boeck20ismir}, we use the %provided test-splits of 
GTZAN \cite{tzanetakis2002musical, marchand2015swing}, Giantsteps \cite{knees2015two, schreiber2018crowdsourced} and ACM-Mirum \cite{peeters2012perceptual, percival2014streamlined} datasets for model evaluation. 

To be comparable to different baseline approaches we use \emph{Accuracy 1} (\emph{Acc1}) and \emph{Accuracy 2} (\emph{Acc2}) \cite{gouyon2006experimental} as evaluation metrics. 
\emph{Acc1} considers a prediction as correct if it falls within a $4\%$ tolerance window of the ground truth tempo. Similarly, \emph{Acc2} accounts for octave errors by also allowing half, double, third and three times the given ground truth tempo with $4\%$ tolerance. 

\subsection{Baselines}
We consider three main baselines from the literature -- a convolutional neural network (Schreiber) \cite{schreiber18ismir}, a jointly trained tempo, beat and downbeat estimation approach (B\"{o}ck) \cite{boeck20ismir} as well as a classifier trained on top of embeddings that were learned in a self-supervised fashion using an equivariance constraint (Quinton) \cite{quinton22ismir}. Note once more that all of these baselines used labeled data, in some capacity, to train the actual tempo classification model, whereas our method has never been exposed to a real annotated piece of music. Thus, the goal here is not specifically to show that we can outperform the current state-of-the-art, but rather to explore how comparable a strictly self-supervised approach can be to reference approaches which include at least some level of supervised learning.\footnote{Due to the lack of an evaluation on real-world data we omit \cite{morais2023tempo} as a baseline.}

In addition, we include two tempo estimation algorithms (Gkiokas~\cite{DBLP:conf/icassp/GkiokasKCS12} and Percival~\cite{percival2014streamlined}) whose approaches depend more on domain knowledge and signal processing than machine-learning. In this sense, both are essentially ``unsupervised''.
Hence, by comparing these, we expect to observe how the proposed method can perform in a similar condition where tempo labels are unavailable.\footnote{We report the results from \cite{bock2019multi} as the original works do not provide results on the datasets of interest.}

Finally, we propose an additional internal baseline in the form of a $1$-nearest neighbor (NN) classifier, which computes the cosine similarities between all reference and target MULE embeddings and picks the tempo of the closest reference as the final prediction. Like our proposed approach, this method does not rely on human labeled data.

\section{Results}
\label{sec:results}
Our results are summarized in Table \ref{table:results}.
While our SDNet\textsubscript{argmax} approach falls some way behind the domain-knowledge-based detection algorithms and supervised baselines for \emph{Acc1}, we observe %very 
competitive \emph{Acc2} scores. This indicates that our approach has a strong notion of tempo, but fails to identify the correct octave, which is a common problem in tempo estimation \cite{gouyon2006experimental,schreiber2017post} and often even ambiguous to humans \cite{mckinney06mp}. Additionally, it is worth highlighting the nearest neighbor approach which also serves as a strong baseline in terms of \emph{Acc2}. This suggests that the MULE embeddings themselves already encode tempo information, even though they were not trained to specifically do so compared to \cite{quinton22ismir}. While this baseline yields only marginally lower scores for \emph{Acc2}, we observe higher \emph{Acc1} across all datasets for our SDNet\textsubscript{argmax} approach.

\begin{table}[t!]
	\centering
	\footnotesize
	\begin{tabular}{@{}lcccccccccc@{}}
		\multirow{2}{*}{\textbf{}} & \multicolumn{2}{c}{GTZAN} & \multicolumn{2}{c}{ACM-Mirum} & \multicolumn{2}{c}{Giantsteps} \\
		& Acc1 & Acc2 & Acc1 & Acc2 & Acc1 & Acc2 \\
		\toprule[1.1pt]
		
		MULE $1$-NN  & 42.4 & 87.4 & 37.0 & 92.1 & 35.9 & 96.1 \\ \midrule
		SDNet\textsubscript{argmax}                              & 53.1 & 89.4  & 57.9  & 93.4  & 47.4 & 97.6 \\
		SDNet\textsubscript{corrected}     & 65.1 & 90.0  & 64.3  & 95.6  & 72.8 & 97.9 \\\midrule[1.1pt]
		Gkiokas \cite{DBLP:conf/icassp/GkiokasKCS12, bock2019multi} & 65.1 & 93.1 & 72.5 & 97.9 & 72.1 & 92.2 \\
		Percival \cite{percival2014streamlined, bock2019multi} & 65.8 & 92.4 & 73.3 & 97.2 & 50.6 & 95.6 \\\midrule[1.1pt]
		Quinton \cite{quinton22ismir}      & 74.1 & 91.9 & 74.7 & 96.5 & 47.0 & 88.6 \\
		Schreiber \cite{schreiber18ismir}  & 76.9 & 92.6 & 78.1 & 97.6 & 82.1 & 97.1 \\
		B\"{o}ck \cite{boeck20ismir}       & 83.0 & 95.0 & 84.1 & 99.0 & 87.0 & 96.5 \\
		\bottomrule[1.1pt]
	\end{tabular}  
	\caption{Overview of tempo estimation accuracy. For Quinton we choose the model configuration with $r_f=0.2$.}
	\label{table:results}
\end{table}

\begin{figure*}[t!]
	\centerline{
		\includegraphics[width=0.9\textwidth]{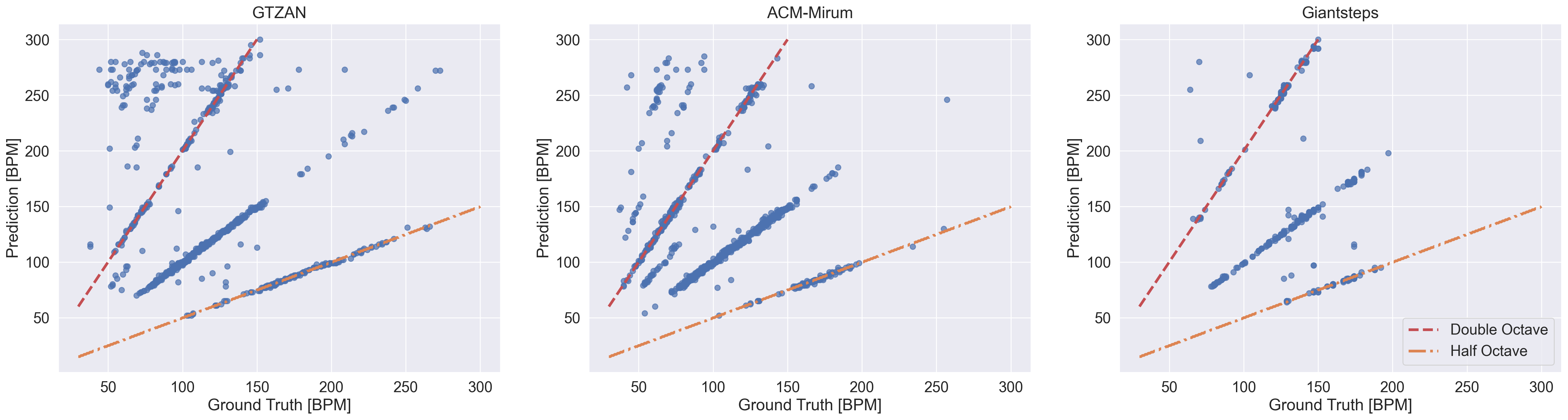}
	}
	\caption{Visualization of predicted vs. ground truth tempo for GTZAN, ACM-Mirum and Giantsteps using SDNet\textsubscript{argmax}. As indicated by the dashed lines, the model makes multiple \emph{octave} errors.}
	\label{fig:output_analysis}
\end{figure*}

\subsection{Error Analysis}
To shed more light on the errors of our model, Figure \ref{fig:output_analysis} visualizes the predictions compared to the ground truth annotations across all three test datasets. As already indicated by the high \emph{Acc2} scores, we clearly see that the model struggles with \emph{half} and \emph{double} octaves. Indeed, on GTZAN and ACM-Mirum we even observe octave errors as high as $3$ and $4$ times the annotated tempo. While overall, the model does not make many errors beyond those related to tempo octave on ACM-Mirum and Giantsteps, we see several misclassifications for GTZAN where the predictions do not fall on any obvious diagonal line in the scatter plot. Closer inspection reveals this clustering of poorly estimated examples occurs at slower tempi (with ground truth annotations mostly less than $100$ BPM). Looking beyond global tempo as a potential complicating factor, examination of the labeled genres of these specific excerpts revealed a high proportion of classical music. %Audition 
Listening to these excerpts and inspection of the corresponding beat annotations \cite{marchand2015swing} indicated highly variable tempo. In cases like these, \cite{schreiber2020music}
argues that the $Acc1$ and $Acc2$ scores are not informative and it is rather the \textit{local} tempo which should be predicted and evaluated.

\subsection{Tempo octave adjustment}
\begin{figure}[t]
	\centerline{
		\includegraphics[width=0.9\columnwidth]{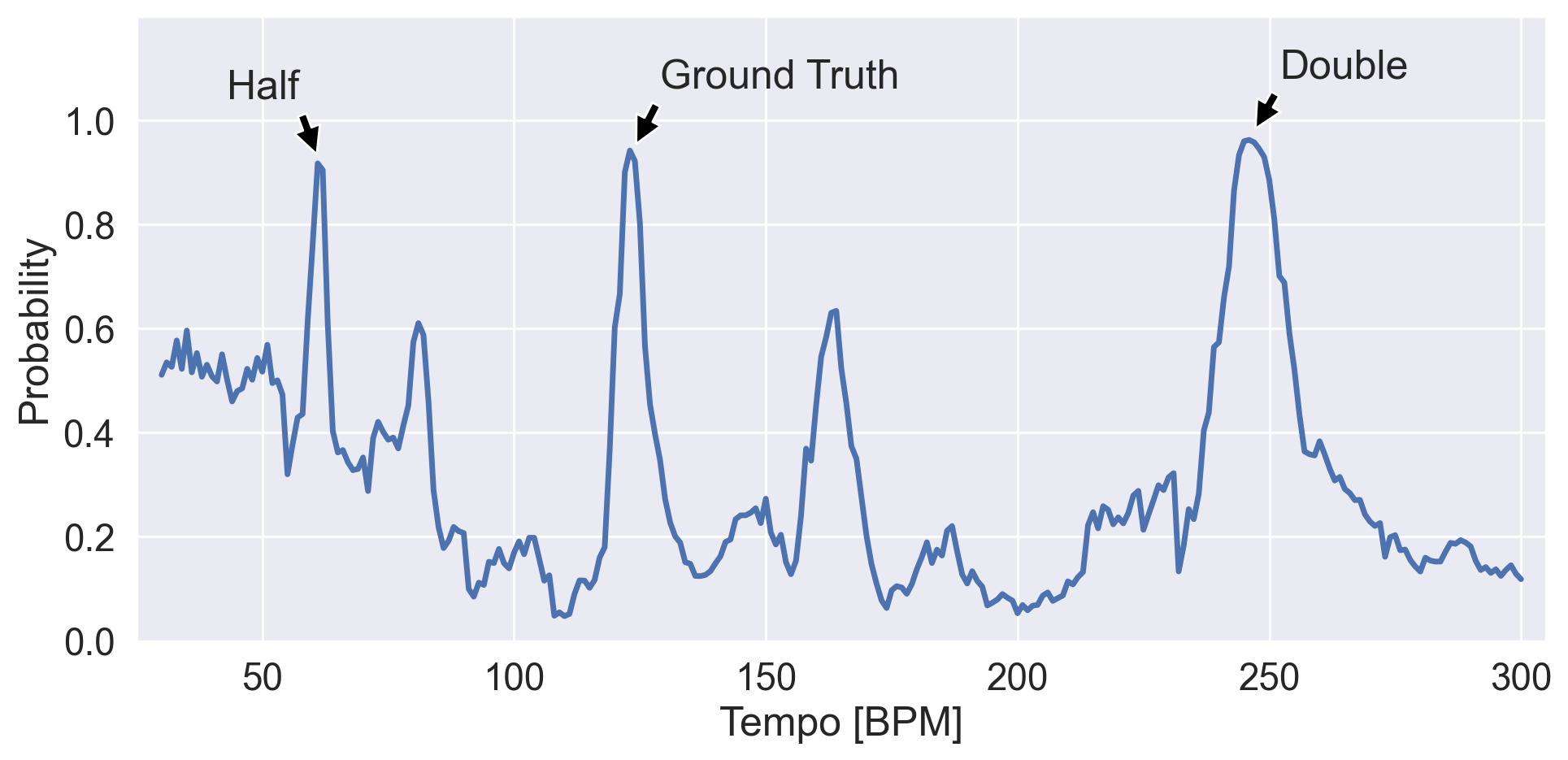}
	}
	\caption{Typical example of an octave error in the GTZAN dataset. We observe high probabilities for the actual annotation as well as the half and double octave. 
	}
	\label{fig:output_probabilities}
\end{figure}

Further analyzing cases of octave errors, we often observe high output probabilities across both \emph{half} and \emph{double} octaves as well as the actual annotated ground truth (cf. Figure \ref{fig:output_probabilities}). This makes it particularly hard to pick a peak as the probabilities tend not to vary significantly. Regardless, using this information alongside the fact that tempo is not uniformly distributed across the entire range of 30 to 300 BPM (as shown for commonly used annotated datasets in \cite{boeck20ismir}), we can gain further insight into our method using a simple post-hoc correction strategy that searches for the three highest tempo peaks, checks if they can be considered half and double octaves, and if this is the case, picks the middle octave as the most likely tempo, otherwise we fall back to the \emph{argmax} solution. We refer to this as SDNet\textsubscript{corrected} in Table \ref{table:results}. While we recognise that this correction is heuristic in nature, and thus its impact is somewhat dependent on the tempo distribution in the test datasets, we observe considerable improvements in \emph{Acc1} for all datasets, as well as small improvements in \emph{Acc2}.
In comparison to the conventional tempo estimation algorithms, SDNet\textsubscript{corrected} achieves competitive \emph{Acc1} results in two out of three and comparable \emph{Acc2} scores across all datasets.
This suggests that the proposed method can learn the tempo estimation logic comparable to those based on sophisticated domain knowledge, with only a relatively simple learning objective and correction logic. Furthermore this is achievable from a task-agnostic musical audio embedding.

\section{Discussion and Conclusions}
\label{sec:conclusion}
In this work we have proposed a fully self-supervised approach to music tempo estimation. We formulate the task as a binary classification problem which hinges on whether the tempo is the same or different between the input and a set of synthetic tempo references in a general purpose musical embedding space. While we obtain highly competitive performance compared to several baseline approaches under the less strict \emph{Acc2} score, we observe a specific limitation in the ability of our approach to correctly assign the annotated tempo octave, and thus we do not achieve similarly high \emph{Acc1} performance. The inclusion of a peak selection heuristic helped partially close this \emph{Acc1} to \emph{Acc2} gap, and while this relies on domain knowledge we nevertheless consider it important to reiterate that our approach is entirely self-supervised and operates without having ever been exposed to real world musical content with explicit tempo annotations. On this basis, we believe the principal contributions of this work rest in the relative simplicity of the formulation of tempo estimation as binary classification together with the first quantitative evaluation on real world data for a fully self-supervised approach.

Looking beyond the presented content, we identify several promising areas for future work. In particular, we target the improvement of the determination of the ``correct'' (i.e., annotated) tempo octave. To this end, we seek to understand if there is any inherent tempo octave bias in the MULE embedding itself and further investigate whether this is also observable when using different embeddings, as well as exploring the impact of a more timbrally diverse set of synthetic references beyond isolated piano notes. Alternatively, we could relax our fully self-supervised constraint, and follow \cite{quinton22ismir} by adding a supervised fine-tuning stage. 

Regarding the effectiveness of our approach on more challenging music material, e.g., with highly variable tempo, we believe there is strong potential to reformulate our approach as binary classification through time and estimate the local tempo. Albeit more computationally intensive, this could be implemented by replacing the \emph{track-level} embeddings of the unlabeled target track with the \emph{excerpt-level} embeddings in the prediction stage. 

%\vfill
%\pagebreak

%\vfill\pagebreak

\bibliographystyle{IEEEbib}
\bibliography{main}

\begin{thebibliography}{10}

\bibitem{schreiber18ismir}
H.~Schreiber and M.~M{\"u}ller,
\newblock ``A single-step approach to musical tempo estimation using a
  convolutional neural network,''
\newblock in {\em Proc. of the 19th Int. Society for Music Information
  Retrieval Conf.}, 2018, pp. 100--105.

\bibitem{boeck20ismir}
S.~B\"{o}ck and M.~E.~P. Davies,
\newblock ``Deconstruct, analyse, reconstruct: How to improve tempo, beat, and
  downbeat estimation,''
\newblock in {\em Proc. of the 21st Int. Society for Music Information
  Retrieval Conf.}, 2020, pp. 574--582.

\bibitem{quinton22ismir}
E.~Quinton,
\newblock ``Equivariant self-supervision for musical tempo estimation,''
\newblock in {\em Proc. of the 23rd Int. Society for Music Information
  Retrieval Conf.}, 2022, pp. 84--92.

\bibitem{morais2023tempo}
G.~Morais, M.~E.~P. Davies, M.~Queiroz, and M.~Fuentes,
\newblock ``Tempo vs. pitch: understanding self-supervised tempo estimation,''
\newblock in {\em Proc. of the IEEE Int. Conf. on Acoustics, Speech and Signal
  Processing}, 2023, pp. 1--5.

\bibitem{schreiber2020music}
H.~Schreiber, J.~Urbano, and M.~M{\"u}ller,
\newblock ``{Music Tempo Estimation: Are we done yet?},''
\newblock {\em Trans. of the Int. Society for Music Information Retrieval},
  vol. 3, no. 1, 2020.

\bibitem{gouyon2006experimental}
F.~Gouyon, A.~Klapuri, S.~Dixon, M.~Alonso, G.~Tzanetakis, C.~Uhle, and P.Cano,
\newblock ``An experimental comparison of audio tempo induction algorithms,''
\newblock {\em IEEE Trans. on Audio, Speech, and Language Processing}, vol. 14,
  no. 5, pp. 1832--1844, 2006.

\bibitem{mckinney07jnmr}
M.~F. McKinney, D.~Moelants, M.~E.~P. Davies, and A.~Klapuri,
\newblock ``Evaluation of audio beat tracking and music tempo extraction
  algorithms,''
\newblock {\em Journal of New Music Research}, vol. 36, no. 1, pp. 1--16, 2007.

\bibitem{gfeller2020spice}
B.~Gfeller, C.~Frank, D.~Roblek, M.~Sharifi, M.~Tagliasacchi, and
  M.~Velimirovi{\'c},
\newblock ``{SPICE}: Self-supervised pitch estimation,''
\newblock {\em IEEE/ACM Trans. on Audio, Speech, and Language Processing}, vol.
  28, pp. 1118--1128, 2020.

\bibitem{grosche11taslp}
P.~Grosche and M.~M{\"u}ller,
\newblock ``Extracting predominant local pulse information from music
  recordings,''
\newblock {\em IEEE Transactions on Audio, Speech, and Language Processing},
  vol. 19, no. 6, pp. 1688--1701, 2011.

\bibitem{mccallum22ismir}
M.~C. McCallum, F.~Korzeniowski, S.~Oramas, F.~Gouyon, and A.~Ehmann,
\newblock ``Supervised and unsupervised learning of audio representations for
  music understanding,''
\newblock in {\em Proc. of the 23rd Int. Society for Music Information
  Retrieval Conf.}, 2022, pp. 256--263.

\bibitem{li2023mert}
Y.~Li, R.~Yuan, G.~Zhang, Y.~Ma, X.~Chen, H.~Yin, C.~Lin, A.~Ragni, E.~Benetos,
  N.~Gyenge, R.~Dannenberg, R.~Liu, W.~Chen, G.~Xia, Y.~Shi, W.~Huang, Y.~Guo,
  and J.~Fu,
\newblock ``{MERT: Acoustic Music Understanding Model with Large-Scale
  Self-supervised Training},''
\newblock {\em arXiv preprint arXiv:2306.00107}, 2023.

\bibitem{mccallum24icassp}
M.~C. McCallum, M.~E.~P. Davies, F.~Henkel, J.~Kim, and S.~Sandberg,
\newblock ``On the effect of data-augmentation on local embedding properties in
  the contrastive learning of music audio representations,''
\newblock in {\em Proc. of the {IEEE} Int. Conf. on Acoustics, Speech, and
  Signal Processing}, 2024.

\bibitem{mccallum24icassp2}
M.~C. McCallum, F.~Henkel, J.~Kim, S.~Sandberg, and M.~E.~P. Davies,
\newblock ``Similar but faster: manipulation of tempo in music audio embeddings
  for tempo prediction and search,''
\newblock in {\em Proc. of the IEEE Int. Conf. on Acoustics, Speech and Signal
  Processing}, 2024.

\bibitem{chen2020simclr}
T.~Chen, S.~Kornblith, M.~Norouzi, and G.~Hinton,
\newblock ``A simple framework for contrastive learning of visual
  representations,''
\newblock in {\em Proc. of the 37th Int. Conf. on Machine Learning}, 2020, pp.
  1597--1607.

\bibitem{nair2010rectified}
V.~Nair and G.~Hinton,
\newblock ``Rectified linear units improve restricted boltzmann machines,''
\newblock in {\em Proc. of the 27th Int. Conf. on Machine Learning}, 2010, pp.
  807--814.

\bibitem{kingma14adam}
D.~P. Kingma and J.~Ba,
\newblock ``Adam: {A} method for stochastic optimization,''
\newblock in {\em Proc. of the 3rd Int. Conf. on Learning Representations},
  2015.

\bibitem{LoshchilovH17_SGDR_ICLR}
I.~Loshchilov and F.~Hutter,
\newblock ``{{SGDR:} Stochastic Gradient Descent with Warm Restarts},''
\newblock in {\em Proc. of the 5th Int. Conf. on Learning Representations},
  2017.

\bibitem{gemmeke2017audioset}
J.~F. Gemmeke, D.~P.~W. Ellis, D.~Freedman, A.~Jansen, W.~Lawrence, R.~C.
  Moore, M.~Plakal, and M.~Ritter,
\newblock ``{Audio Set: An ontology and human-labeled dataset for audio
  events},''
\newblock in {\em Proc. of the {IEEE} Int. Conf. on Acoustics, Speech, and
  Signal Processing}, 2017, pp. 776--780.

\bibitem{bertinMahieux2011msd}
T.~Bertin-Mahieux, D.~P.W. Ellis, B.~Whitman, and P.~Lamere,
\newblock ``The million song dataset,''
\newblock in {\em Proc. of the 12th Int. Conf. on Music Information Retrieval},
  2011, pp. 591--596.

\bibitem{tzanetakis2002musical}
G.~Tzanetakis and P.~Cook,
\newblock ``Musical genre classification of audio signals,''
\newblock {\em IEEE Trans. on Speech and Audio processing}, vol. 10, no. 5, pp.
  293--302, 2002.

\bibitem{marchand2015swing}
U.~Marchand and G.~Peeters,
\newblock ``Swing ratio estimation,''
\newblock in {\em Proc. of the 18th Int. Conf. on Digital Audio Effects}, 2015,
  pp. 423--428.

\bibitem{knees2015two}
P.~Knees, {\'A}.~Faraldo, H.~Boyer, R.~Vogl, S.~B{\"o}ck,
  FlorianH{\"o}rschl{\"a}ger, and Mickael~Le Goff,
\newblock ``Two data sets for tempo estimation and key detection in electronic
  dance music annotated from user corrections,''
\newblock in {\em Proc. of the 16th Int. Society for Music Information
  Retrieval Conf.}, 2015, pp. 364--370.

\bibitem{schreiber2018crowdsourced}
H.~Schreiber and M.~M{\"u}ller,
\newblock ``A crowdsourced experiment for tempo estimation of electronic dance
  music.,''
\newblock in {\em Proc. of the 19th Int. Society for Music Information
  Retrieval Conf.}, 2018, pp. 409--415.

\bibitem{peeters2012perceptual}
G.~Peeters and J.~Flocon-Cholet,
\newblock ``Perceptual tempo estimation using {GMM}-regression,''
\newblock in {\em Proc. of the 2nd Int. ACM workshop on Music Information
  Retrieval with user-centered and multimodal strategies}, 2012, pp. 45--50.

\bibitem{percival2014streamlined}
G.~Percival and G.~Tzanetakis,
\newblock ``Streamlined tempo estimation based on autocorrelation and
  cross-correlation with pulses,''
\newblock {\em IEEE/ACM Trans. on Audio, Speech, and Language Processing}, vol.
  22, no. 12, pp. 1765--1776, 2014.

\bibitem{DBLP:conf/icassp/GkiokasKCS12}
A.~Gkiokas, V.~Katsouros, G.~Carayannis, and T.~Stafylakis,
\newblock ``Music tempo estimation and beat tracking by applying source
  separation and metrical relations,''
\newblock in {\em Proc. of the {IEEE} Int. Conf. on Acoustics, Speech and
  Signal Processing}, 2012, pp. 421--424.

\bibitem{bock2019multi}
S.~B{\"o}ck, M.~E.~P. Davies, and P.~Knees,
\newblock ``Multi-task learning of tempo and beat: Learning one to improve the
  other.,''
\newblock in {\em Proc. of the 20th Int. Society for Music Information
  Retrieval Conf.}, 2019, pp. 486--493.

\bibitem{schreiber2017post}
H.~Schreiber and M.~M{\"u}ller,
\newblock ``A post-processing procedure for improving music tempo estimates
  using supervised learning.,''
\newblock in {\em Proc. of the 18th Int. Society for Music Information
  Retrieval Conf.}, 2017, pp. 235--242.

\bibitem{mckinney06mp}
M.~F. McKinney and D.~Moelants,
\newblock ``Ambiguity in tempo perception: What draws listeners to different
  metrical levels?,''
\newblock {\em Music Perception: An Interdisciplinary Journal}, vol. 24, no. 2,
  pp. 155--166, 2006.

\end{thebibliography}
\end{document}